\pdfoutput=1
\documentclass[prl,letterpaper,twocolumn,superscriptaddress,showpacs,aps]{revtex4}
\usepackage{graphicx}
\usepackage{amssymb}
\usepackage{amsmath}
\usepackage{amsfonts}
\usepackage{dcolumn}
\usepackage{ogonek}

\begin{document}
\title{New Precise Measurement of the Pion Weak Form Factors in {\boldmath
    $\pi^+\to {\rm e}^+\nu\gamma$} Decay} 

\newcommand*{\uva}{Department of Physics, University of Virginia,
                   Charlottesville, VA 22904-4714, USA}
\newcommand*{\psii}{Paul Scherrer Institut, Villigen PSI, CH-5232,
                    Switzerland}
\newcommand*{\dubna}{Joint Institute for Nuclear Research, RU-141980
                     Dubna, Russia}
\newcommand*{\swierk}{Instytut Problem\'ow J\k{a}drowych im.\
                      Andrzeja So{\l}tana, PL-05-400 \'Swierk, Poland}
\newcommand*{\tbilisi}{Institute for High Energy Physics, Tbilisi
                       State University, GUS-380086 Tbilisi, Georgia}
\newcommand*{\asu}{Department of Physics and Astronomy, Arizona
                   State University, Tempe, AZ 85287-1504, USA}
\newcommand*{\irb}{Institut ``Rudjer Bo\v{s}kovi\'c,'' HR-10000 Zagreb,
                   Croatia}
\newcommand*{\zurich}{Physik-Institut, Universit\"at Z\"urich, CH-8057 Z\"urich, Switzerland}

\affiliation{\uva}
\affiliation{\dubna}
\affiliation{\psii}
\affiliation{\irb}
\affiliation{\swierk}
\affiliation{\tbilisi}
\affiliation{\zurich}

\author{M.~Bychkov}\email[Corresponding author: ]
                   {mab3ed@virginia.edu.}\affiliation{\uva}
\author{D.~Po\v{c}ani\'c}\email[Corresponding author: ]
                   {pocanic@virginia.edu.}\affiliation{\uva}
\author{B.~A.~VanDevender}\affiliation{\uva}
\author{V.~A.~Baranov}\affiliation{\dubna}
\author{W.~Bertl}\affiliation{\psii}
\author{Yu.~M.~Bystritsky}\affiliation{\dubna}
\author{E.~Frle\v{z}}\affiliation{\uva}
\author{V.~A.~Kalinnikov}\affiliation{\dubna}
\author{N.~V.~Khomutov}\affiliation{\dubna}
\author{A.~S.~Korenchenko}\affiliation{\dubna}
\author{S.~M.~Korenchenko}\affiliation{\dubna}
\author{M.~Korolija}\affiliation{\irb}
\author{T.~Kozlowski}\affiliation{\swierk}
\author{N.~P.~Kravchuk}\affiliation{\dubna}
\author{N.~A.~Kuchinsky}\affiliation{\dubna}
\author{W.~Li}\affiliation{\uva}
\author{D.~Mekterovi\'c}\affiliation{\irb}
\author{D.~Mzhavia}\affiliation{\dubna}\affiliation{\tbilisi}
\author{S.~Ritt}\affiliation{\psii}
\author{P.~Robmann}\affiliation{\zurich}
\author{O.~A.~Rondon-Aramayo}\affiliation{\uva}
\author{A.~M.~Rozhdestvensky}\affiliation{\dubna}
\author{T.~Sakhelashvili}\affiliation{\psii}
\author{S.~Scheu}\affiliation{\zurich}
\author{U.~Straumann}\affiliation{\zurich}
\author{I.~Supek}\affiliation{\irb}
\author{Z.~Tsamalaidze}\affiliation{\dubna}\affiliation{\tbilisi}
\author{A.~van~der~Schaaf}\affiliation{\zurich}
\author{E.~P.~Velicheva}\affiliation{\dubna}
\author{V.~P.~Volnykh}\affiliation{\dubna}
\author{Y.~Wang}\affiliation{\uva}
\author{H.-P.~Wirtz}\affiliation{\psii}

\date{9 Apr 2008, revised 24 June 2009}
\begin{abstract}
We have measured the $\pi^+\to {\rm e}^+\nu\gamma$ branching ratio
over a wide region of phase space, based on a total of 65,460 events
acquired using the PIBETA detector.  Minimum-$\chi^2$ fits to the
measured $(E_{e^+},E_\gamma)$ energy distributions result in the weak
form factor value of $F_A=0.0119(1)$ with a fixed value of
$F_V=0.0259$. An unconstrained fit yields $F_V=0.0258(17)$ and
$F_A=0.0117(17)$. In addition, we have measured $a=0.10(6)$
for the dependence of $F_V$ on $q^2$, the ${\rm e}^{+}\nu$ pair
invariant mass squared, parametrized as $F_V(q^2)=F_V(0)(1+a\cdot
q^2)$. The branching ratio for the kinematic region $E_\gamma >
10\,$MeV and $\theta_{{\rm e^+}\gamma} > 40^\circ $ is measured to be
$B^{\rm exp}=73.86(54) \times 10^{-8}$. Earlier deviations we reported
in the high-$E_\gamma$/low-$E_{{\rm e}^+}$ kinematic region are
resolved, and we find full compatibility with CVC and standard
$V$$-$$A$ calculations without a tensor term.  We also derive new
values for the pion polarizability, $\alpha_E = \rm 2.78(10) \times
10^{-4}\,fm^3$, and neutral pion lifetime, $\tau_{\pi 0} = (8.5 \pm
1.1) \times 10^{-17}\,$s.

\end{abstract}
\maketitle

Radiative pion decay $\pi^+\to {\rm e}^+\nu\gamma$ (also denoted
$\pi_{{\rm e2}\gamma}$), where $\gamma$ is a real photon, offers the
best means to study pion form factors at zero momentum transfer.  Pion
form factors provide critical input to low-energy effective theories
of the strong interaction, such as chiral perturbation theory (ChPT).
Previous experimental studies of the $\pi_{{\rm e}2\gamma}$ decay were
constrained by relatively low event statistics and limited kinematic
coverage, leaving room for speculative interpretations including an
anomalously large tensor interaction term.  In this Letter we report
precise new results of a combined analysis of data recorded with the
PIBETA detector~\cite{Frl03a}, removing the limitations and
ambiguities found in previous work.

In the standard description, the $\pi_{{\rm e2}\gamma}$ decay
amplitude consists of the inner bremsstrahlung (IB),
structure-dependent (SD$^\pm$), and interference terms.  The SD terms
are parametrized by vector ($F_V$) and axial vector ($F_A$) form
factors (see Ref.~\cite{Bry82} for a review).  The conserved vector
current (CVC) hypothesis relates $F_V$ to the $\pi^0$
lifetime~\cite{Vak58,Ger55,Fey58}, yielding
$F_V=0.0259(9)$~\cite{PDG08}.  ChPT
calculations~\cite{Hol86,Bij97,Gen03, Pic08} give a value for $F_A$ in
the range 0.010--0.012.

Due to the above noted limitations, authors of previous experimental
studies fixed $F_V$ at the CVC-predicted value and evaluated $F_A$,
with resulting relative uncertainties ranging from $\pm$12\,\% to
$\pm$56\,\%~\cite{Dep63,Ste78,Bay86,Pii86,Dom88,Bol90}.  None of these
measurements was sensitive to the form factor dependence on the ${\rm
e}^{+}\nu$ pair invariant mass.  Because of inconsistencies in the
available data, a relatively large non-($V$$-$$A$) tensor contribution
to the SD terms was considered (see Ref.~\cite{Chi04} and references
therein).  Our first measurement~\cite{Frl04} reported a significant
improvement in the accuracy of $F_A$, but also noted a substantial
deficit of observed events in the high-$E_\gamma$/low-$E_{{\rm e}^+}$
kinematic region, still leaving room for non-($V$$-$$A$) admixtures.

We performed measurements at the $\pi E1$ beam line at the Paul
Scherrer Institute (PSI), Switzerland, using a stopped $\pi^+$ beam
and the PIBETA large-acceptance CsI calorimeter with central
tracking~\cite{Frl03a}.  A total of over $2.0\times 10^{13}$ $\pi^+$'s
stopped in the active target during the two measuring periods were
used in the analysis. The first data set (1999--2001) was collected
during the $\pi^+\to \pi^0{\rm e}^+\nu$ ($\pi_\beta$) branching ratio
measurement~\cite{Poc04} with $\sim 8\times 10^{5}$ pion stops per
second.  A second sample was recorded in 2004 with an eight times
lower stopping rate.  Details of the detector
design, performance, and our experimental methods are presented in
Refs.~\cite{Frl03a,Poc04,Frl04,pb_web}.

The PIBETA detector measured shower energies, and directions of the
positron and photon, thus over-determining the kinematics of the final
three-body state.  Key for the $\pi_{\rm e2\gamma}$ measurement were
the one-arm and two-arm high-threshold triggers, 1HT and 2HT,
respectively, requiring that at least one (1HT) or two (2HT)
showers register above an effective high threshold of $\sim
54\,$MeV deposited in the calorimeter, thus reducing a copious e$^+$
background from the $\pi$-$\mu$-e decay chain with a continuous energy
spectrum ending at $\sim 52.8\,$MeV.  In the 99-01 data set the 1HT
trigger was prescaled 16- or 8-fold.  Our $\pi_{\rm e2\gamma}$ data
set covers three non-overlapping kinematic regions given in
Tab.~\ref{tab:0}; region $I$ data were acquired with the 2HT trigger,
while the 1HT trigger mapped out regions $II$ and $III$.
In all three regions the $\pi_{\rm e2\gamma}$ decay is strongly
dominated by the SD$^+$ term proportional to $(F_V+F_A)^2$
\cite{Bry82}.  Candidate $\pi_{{\rm e}2\gamma}$ events exhibit one
neutral shower in the calorimeter coincident in time with a positron
track. For the events with more than one available combination, i.e.,
more than one neutral shower or positron track, the pair most nearly
coincident in time was chosen.  Events
incompatible with the $\pi_{{\rm e2}\gamma}$ kinematics were rejected in
the final data sample.

\begin{table} [t]
\caption{Total energy ranges for kinematic regions used in measurement
  ($I$--$III$, in measured MeV$_{\rm exp}$) and branching ratio
  evaluation ($A$--$O$, in physical MeV).  For regions $B$, $C$, and
  $O$, the e$^+$-$\gamma$ relative angle is restricted to
  $\theta_{{\rm e}^+\gamma} >40^\circ$; this condition is
  automatically satisfied for regions $I$-$III$ and $A$. }
\label{tab:0}
\begin{ruledtabular}
\begin{tabular}{lcc|lcc}
 \multicolumn{3}{c|}{Measurement} & \multicolumn{3}{c}{B.R. evaluation} \\
 Reg. & $E^{\rm exp}_{\rm e+}$ & $E^{\rm exp}_\gamma$ 
                                 & Reg. & $E_{\rm e+}$ & $E_\gamma$\\
\hline
  $I$   &  $> 51.7$    &  $>51.7$   &   $A$   &  $>50 $  &  $>50$  \\  
  $II$  & $20 - 51.7$    &  $>55.6$ &   $B$   &  $>10 $  &  $>50$  \\  
  $III$ &  $>55.6$  & $20 - 51.7$   &   $C$   &  $>50 $  &  $>10$  \\  
        &           &               &   $O$   &  $>m_{\rm e}$ & $>10$ \\
\end{tabular}
\end{ruledtabular}
\end{table}

Due to the nature of the detector resolution function, the
experimental kinematic region contains a fraction of events shifted
from a broader kinematic range.  We therefore used the reconstructed
events in regions $I$, $II$, and $III$ to evaluate the branching
ratios $B_{\pi{\rm e}2\gamma}^{ \rm exp}$ for the larger regions $A$,
$B$, and $C$, respectively, and $O$, the combination of all three 
regions, by means of Monte Carlo (MC) extrapolation.
On the other hand, we calculated the theoretical branching
fractions $B_{\pi{\rm e}2\gamma}^{ \rm the}$ for the same regions
numerically using only the input parameters ($F_V, F_A, a$), as
described below, and the standard description of the decay, including
radiative corrections~\cite{Bys04}.

Accidental background was dominated by positrons from $\pi_{{\rm e2}}$
and $\pi$-$\mu$-$\rm e$ decays accompanied by an unrelated neutral
shower.  It was corrected for by subtracting histograms of observables
projected using the out-of-time cut $\Delta t_{\rm out} :
(5\,{\rm ns}< \left| t_{\rm e^+}-t_\gamma \right|
<10\,{\rm ns})$ from the in-time histograms projected via the cut
$\Delta t_{\rm in}:
(\vert t_{\rm e^+}-t_\gamma\vert \le \rm \,5\,ns)$.
The 2004 data sample with lower stopped beam
intensity provided an improvement of the signal to accidental
background ratios from 1.7:1 to 10:1 in region $II$ and 4.4:1 to 30:1
in region $III$.

Non-accidental background sources were predominately $\pi_\beta$ events
for which one $\pi^0$ decay photon converts in the target, producing a
charged track in the detector. Having measured the net yield of the
$\pi_\beta$ decay events in our detector, we have used a MC simulation
to calculate the fraction of $\pi_\beta$ decay events misidentified as
$\pi_{{\rm e}2\gamma}$.  This class of background events contributes
$\sim 6.0\,\%$ of the signal in the kinematic region $I$ and
negligible amounts in regions $II$ and $III$.  The $\pi_\beta$
contamination was subtracted in the calculation of $\pi_{{\rm
e}2\gamma}$ yields.

In order to evaluate the branching ratio of the $\pi_{{\rm e}2\gamma}$
decay in any given kinematic region, for normalization we used the
$\pi_{{\rm e}2}$ events, recorded in parallel via the one-arm HT
calorimeter trigger.  This procedure assures the cancellation of the
imprecisely known 
total number of stopped $\pi^+$'s, and
$\epsilon_{\rm trac}^{\rm e}$, the combined tracking efficiency of $\rm
e^+$'s.  No statistically significant $\rm e^+$ energy dependence was
observed for $\epsilon_{\rm trac}^{\rm e}$ \cite{VanD05}.  Thus, the
experimental branching ratio for the $\pi^+\to {\rm e}^+\nu\gamma$
decay can be calculated from the expression
\begin{equation}
   B^{\rm exp}_{\pi {\rm e}2\gamma} = B_{\pi {\rm e}2}
     \frac{N_{\pi {\rm e}2\gamma}}
            {N_{\pi {\rm e}2}\,p_{\pi {\rm e}2}}\,
     \frac{A_{\pi {\rm e}2}}{A_{\pi {\rm e}2\gamma}}\,,\label{eq:B_exp}
\end{equation}
where $N_{\pi {\rm e}2\gamma}$ ($N_{\pi {\rm e}2}$) is the number of
the detected $\pi_{{\rm e2}\gamma}$ ($\pi_{\rm e2}$) events, $A_{\pi
{\rm e}2\gamma}$ ($A_{\pi {\rm e}2}$) is the experimental acceptance
for the given decay type incorporating the appropriate cuts, and
$B_{\pi {\rm e}2}=1.2352(1)\times 10^{-4}$ is the theoretical
$\pi_{\rm e2}$ branching ratio~\cite{Mar93,Cir07}. The factor $p_{\pi
{\rm e}2}$ denotes the prescaling of the one-arm HT trigger; it 
applied to the 1999--2001 data set only.





\begin{figure} [t] 
\includegraphics[width=0.95\columnwidth]{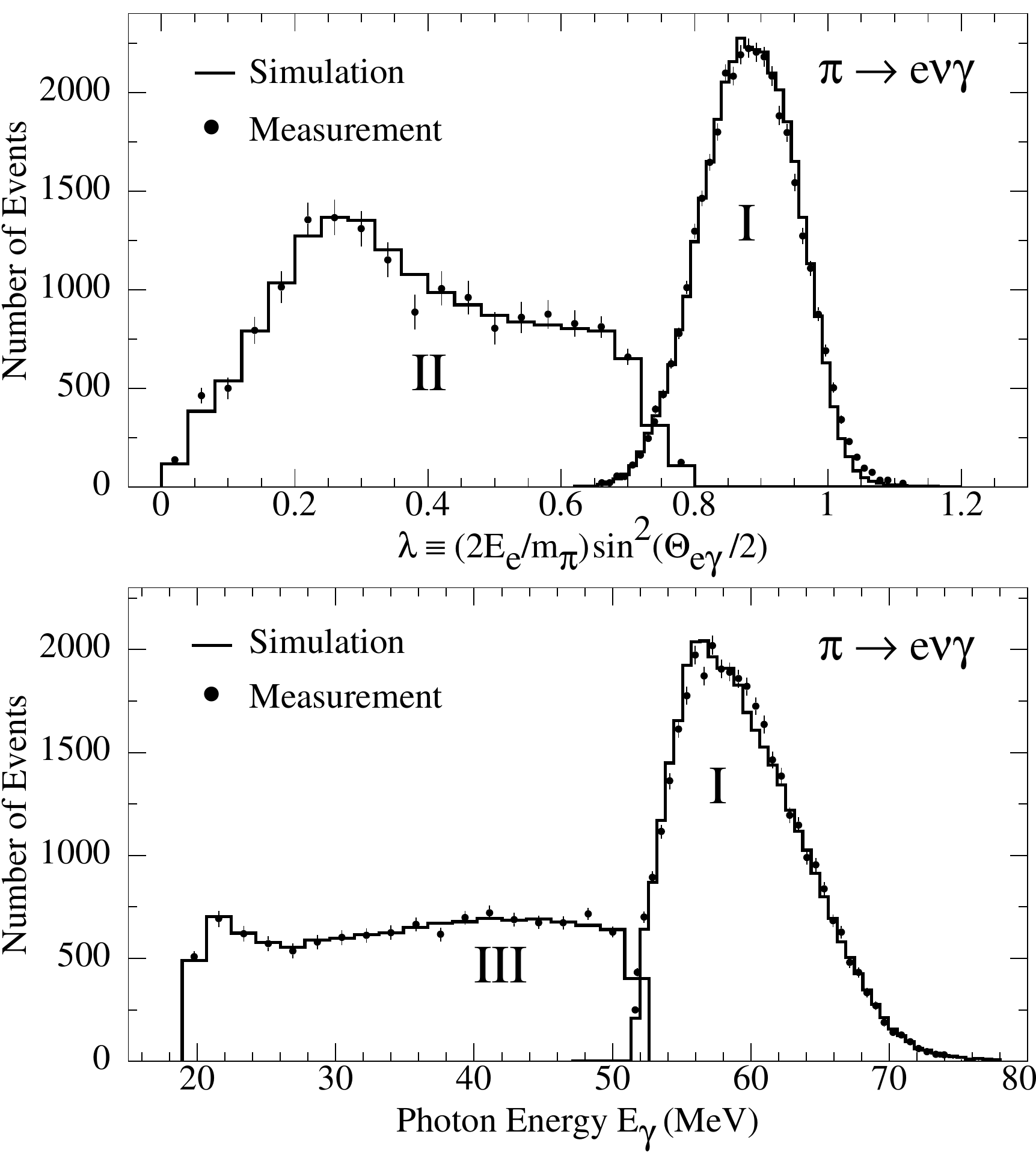}
\caption{Data points: background-subtracted $\pi^+\to {\rm e}^+\nu\gamma$
distribution of the kinematic variable $\lambda \equiv (2E_{\rm
e}/m_{\pi})\sin^{2}(\theta_{{\rm e}\gamma}/2)$ for regions $I$ and $II$
(top panel), and of the photon energy for regions $I$ and $III$ (bottom
panel). Solid lines: results of {\tt GEANT3} calculations for the best-fit
values of $F_V$, $F_A$, and $a$. }
\label{fig:p2eg_lam_ab}
\end{figure}

The experimental acceptances were calculated in a {\tt GEANT3} based MC
program~\cite{Bru94}.  This simulation incorporates a precise
description of the PIBETA detector geometry as well as the electronics
response of the experimental setup. 
Figure~\ref{fig:p2eg_lam_ab} demonstrates the match between the MC
simulation and the background-subtracted data in regions $I$, $II$ and
$III$.  
The number $N_{\pi {\rm e2}}$ was extracted from an independent fit of 
the time distribution of the $\pi_{\rm e2}$ event candidates.
Upon imposing appropriate kinematic cuts and performing the
background subtraction procedure we have reconstructed 35,948 $\pm$
194 ($0.54\%$) events in region $I$, 16,246 $\pm$ 331 ($2.0\%$) events
in region $II$ and 13,263 $\pm$ 161 ($1.2\%$) events in region $III$,
where numbers in parentheses are fractional statistical uncertainties
obtained after accidental background subtraction.  The 2004 data
account for only 8\,\% of the total for region $I$, and about 43\,\%
in regions $II$ and $III$.  However, because of higher accidental
background subtraction, the 2004 events in $II$ and $III$
are statistically more significant than the 1999--2001 events.  More
1999--2001 events are included here than in the analysis reported in
Ref.~\cite{Frl04} thanks to a broadening of the acceptance cuts made
possible by refinements in the treatment of the positron energy
deposited in the target.  The acceptance cuts were adjusted so as to
achieve minimal overall statistical uncertainty after background
subtraction.
The dominant sources of the systematic uncertainties are region
dependent and are summarized in Table~\ref{tab:1}.  Total statistical
and systematic uncertainties for each region are summed in quadrature;
they are given in Table~\ref{tab:2}.

\begin{table} [b]
\caption{Dominant sources of systematic and normalization
uncertainties.  They are: precise determination and simulation of the
trigger thresholds, energy calibration for different types of
particles, radiative corrections to the acceptance and number of the
$\pi_{{\rm e}2}$ events.  Remaining sources include uncertainties in
the signal separation from prompt hadronic and $\pi_{\beta}$
backgrounds.}
\label{tab:1}
\begin{ruledtabular}
\begin{tabular}{lccc}
   Source of uncertainty  & \multicolumn{3}{c}{Region} \\

           & $I\,(\%)$ & $II\,(\%)$ & $III\,(\%)$  \\
\hline
MC of trigger threshold & 0.57 & 0.37 & 1.22 \\
Calorimeter energy calibration & 0.45 & 0.38 & 0.60 \\
Corr. to $\pi_{{\rm e}2}$ acceptance & 0.20 & 0.20 & 0.20  \\
Num. of $\pi_{{\rm e}2}$ events & 0.13 & 0.13 & 0.13  \\
All other sources combined      & 0.19 & 0.02 & 0.05 \\ \hline
Total                           & 0.79 & 0.58 & 1.38 \\
\end{tabular}
\end{ruledtabular}
\end{table}

In order to extract the values of $F_V$, $F_A$, and $a$, and calculate the
values of the branching ratios, we minimized the $\chi^2$ sum of
differences between the experimental and calculated differential
branching ratios over all phase space, taking into account
the form factor dependence on the ${\rm e}^{+}\nu$ pair invariant mass
such that $F_V(q^2)=F_V(0)(1+a\cdot q^2)$, $F_A(q^2)=F_A(0)$, and
$q^2=1-(2E_{\gamma}/m_{\pi})$, following the prescription of
Refs.~\cite{Bij97,Gen03,Pic08}, where $a$ is the slope parameter.  The
experimental branching ratios $B^{\rm exp}_{i}$ acquire a form factor
dependence through the acceptances $A_{\pi {\rm e}2\gamma}$ which are
recalculated at every iteration step of the minimization.

Both theoretical and experimental values of the branching ratios are
proportional to $B_{\pi {\rm e2}}$, therefore our $\chi^2$ sum assures
that the values of the extracted parameters are independent of the
value of $B_{\pi {\rm e2}}$.

For comparison with previous work, and to illustrate the sensitivity
of our data, we performed a fit with fixed values of $F_V =
0.0259$~\cite{PDG08}, $a = 0.041$~\cite{Por07}, yielding $F_A=(119 \pm
1)\times 10^{-4}$ for the free parameter, with
$\rm\chi^2_+/n.d.f.=0.96$. 
This result represents a sixteenfold improvement in precision over the
pre-PIBETA world average~\cite{PDG04}.  Table~\ref{tab:2} and
Fig.~\ref{fig:p2eg_lam_ab} show the excellent agreement between the
measured integral and differential branching ratios, and those
calculated for the best-fit parameter values, reflecting the
exceptionally good description of our data by CVC and standard
$V$$-$$A$ theory in all regions of phase space.  The previously
reported anomalous shortage of events in region $B$ \cite{Frl04} is
thus resolved.
In this work the calorimeter energy calibration (CEC) for
$\gamma$-induced showers was uncoupled from that of the $\rm
e^+$-induced showers, resulting in a $\sim$2\,\% difference.  The clean
2004 data set revealed this discrepancy for the first time in the 1HT
$\pi_\beta$ decay event sample.  Due to the limited statistical
significance of such events in the 1999--2001 set, it was not feasible
to evaluate a separate $\gamma$ CEC in our original analysis.  Because
of the extreme sensitivity of region $II$ to the $\gamma$ energy
scale, the effect was significant only in this region.
More details can be found in Ref.~\cite{Bych05} and in a forthcoming
publication. 

\begin{table} [b]
\caption{Best-fit $\pi_{{\rm e2}\gamma}$ branching ratios obtained with
$F_V=0.0259$ (fixed), $a=0.041$ (fixed) and $F_A =
0.0119(1)$ (fit). Experimental uncertainties in parentheses reflect
both the statistical and systematic uncertainties, given
above. Theoretical uncertainties are dominated by the fit uncertainty
in $F_A$.}
\label{tab:2}
\begin{ruledtabular}
\begin{tabular}{cccl}
   Region  & $B_{\rm exp} \times 10^{8}$  & $B_{\rm the} \times 10^{8}$ & $B_{\rm exp}/B_{\rm the}$ \\
\hline
 $A$ & $2.614(21)$ & $2.599(11) $ & $1.006(9)$ \\
 $B$ & $14.46(22)$ & $14.45(2) $  & $1.001(15)$\\
 $C$ & $37.69(46)$ & $37.49(3) $  & $1.005(12)$\\
 $O$ & $73.86(54)$ & $74.11(3) $  & $0.997(7)$\\
\end{tabular}
\end{ruledtabular}
\end{table}

\begin{figure} [ht] 
\includegraphics[width=0.97\columnwidth]{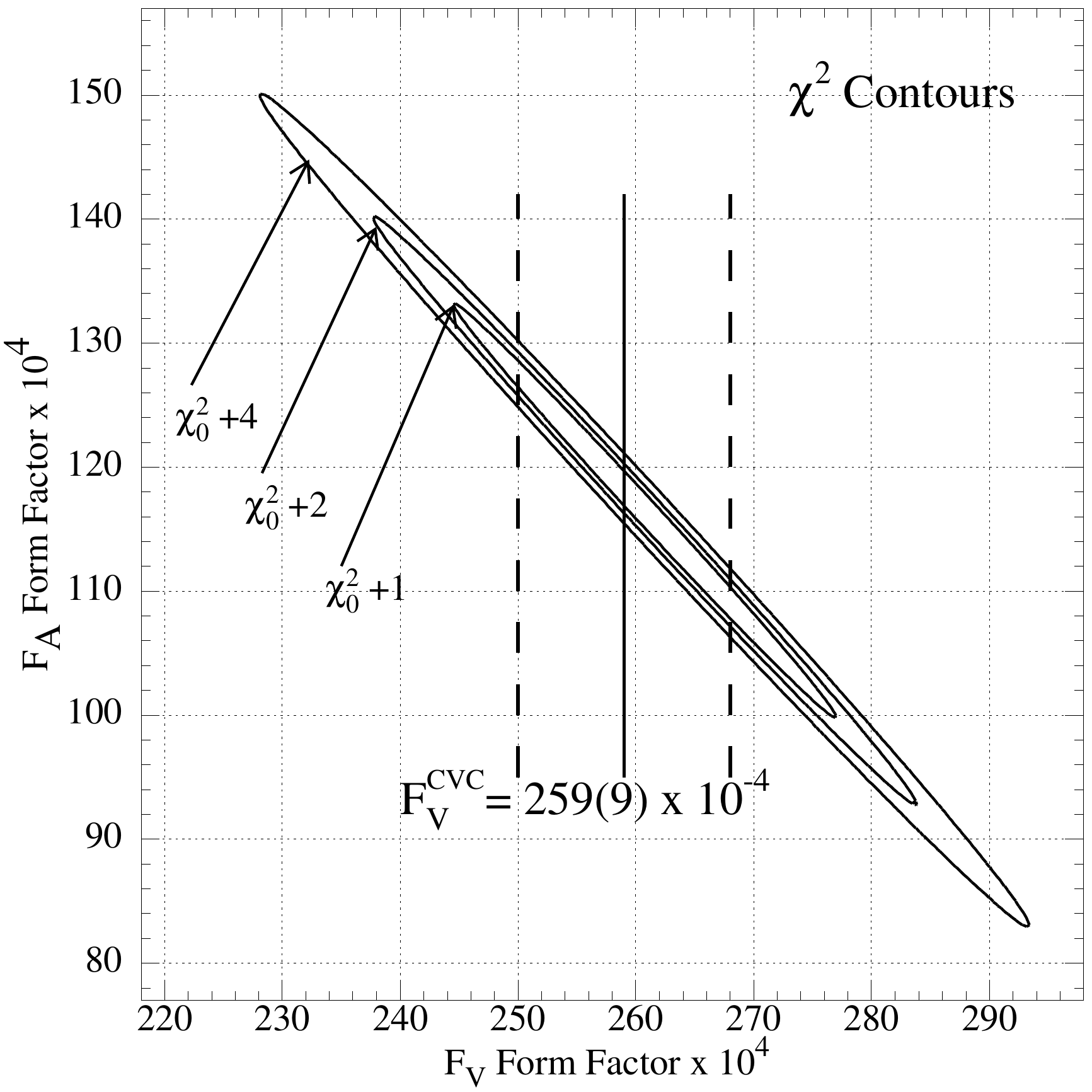}
\caption{Contour plot of loci of constant $\chi^{2}$ for the minimum
  value $\chi_0^2$ plus 1, 2, and 4 units, respectively, in the
  $F_A$-$F_V$ parameter plane, keeping the parameter $a=0.041$.
  The range of the CVC prediction, $F_V = 0.0259(9)$ is indicated.}
\label{fig:fa_fv_cont}
\end{figure}

Because of the strong dominance by the SD$^+$ term, our data are most
sensitive to the combination $F_A+F_V$.  This sensitivity is
illustrated in Fig.~\ref{fig:fa_fv_cont} produced by a two-parameter
($F_A,F_V$) fit.  The best-fit value of $F_A$ has an empirical linear
dependence on the value of $F_V$ parametrized as
\begin{equation}
   F_A=(-1.0286 \cdot F_V + 0.03853) \pm 0.00014 \, .
      \label{eq:fa_fv_fit}
\end{equation}
This expression will remain applicable as the CVC-predicted value of
$F_{V}$ changes thanks to future improved measurements of $\tau_{\pi
0}$, the neutral pion lifetime~\cite{Primex}.

We also examined the question of non-($V$$-$$A$) contributions to the
$\pi_{{\rm e2}\gamma}$ decay by introducing tensor coupling following
the prescription of Ref.~\cite{Chi04}.  The optimal fit yields a
single tensor form factor, $F_T=(-0.6 \pm 2.8)\times 10^{-4}$ , or
$-5.2 \times 10^{-4} < F_T < 4.0 \times 10^{-4}$ with $90\%$
confidence.  Our limits on $F_T$ are in excellent agreement with
Refs.~\cite{Por07,Vol92}, but are two orders of magnitude more
constraining than the results reported by the ISTRA
group~\cite{Bol90,Pob03}.

Finally, in an unconstrained $V$$-$$A$ fit we released all three
parameters $F_A$, $F_V$, and $a$ simultaneously, and obtained
$F_A=0.0117(17)$, $F_V=0.0258(17)$ and $a=0.10(6)$.  These results
are nearly identical to those of the two-parameter $(F_A,F_V)$ fit
presented in Fig.~\ref{fig:fa_fv_cont} and Eq.~(\ref{eq:fa_fv_fit}).

In summary, our results: (i) are in excellent agreement with the CVC
hypothesis prediction for $F_V$, (ii) represent a fourteenfold
improvement in the precision of $F_V$ and a similar improvement in
$F_A$, and (iii) provide the first ever measurement of the charged
pion form factor slope parameter $a$, also found to be consistent with
CVC.  The PEN experiment~\cite{PEN}, currently under way at PSI, will
add substantially to our $\pi_{\rm e2\gamma}$ data set.  Our best-fit
value of $F_A$ agrees well with ChPT calculations, tending to the top
of the reported range~\cite{Hol86,Bij97,Gen03}.  However, a more
precise measurement of $\tau_{\pi 0}$ is needed in order that the
sensitivity of our data, expressed in Eq.~(\ref{eq:fa_fv_fit}), be put
to full use in determining $F_A$.  We use our form factor results to
evaluate the pion polarizability and the ChPT parameter sum
$L_9^r+L_{10}^r$ as follows: using the one-parameter fit we obtain
$\alpha_E = -\beta_M = \rm 2.78(2)_{exp}(10)_{Fv} \times
10^{-4}\,fm^3$, and $L_9^r+L_{10}^r = \rm 0.00145(1)_{exp}(5)_{Fv}$,
where the first uncertainty comes from the fit and the second from the
current CVC-derived value of $F_V$.  Alternatively, we get $\alpha_E
=2.7\left(^{+6}_{-5}\right) \times \rm 10^{-4}\,fm^3$ and
$L_9^r+L_{10}^r = \rm 0.0014\left(^{+3}_{-2}\right)$ based on our
unconstrained fit of $F_A$ and $F_V$.  In addition, we use the latter
fit result and CVC to make an independent determination of the neutral
pion lifetime: $\tau_{\pi0} = (8.5 \pm 1.1) \times 10^{-17}\,$s.

We thank Z.~Hochman for critical help in preparing and carrying out
the measurements, M~.V.~Chizhov for valuable discussions and
theoretical guidance, and J.~Portoles and V.~Mateu for their timely
calculation of the charged pion form factor slope.  The PIBETA
experiment has been supported by the U.S. National Science Foundation,
U.S. Department of Energy, the Paul Scherrer Institute, and
the Russian Foundation for Basic Research.

\bibliography{pienug_v4-13}

\end{document}